\documentclass[12pt]{emulateapj}

\usepackage{amsmath}
\usepackage{rotating}

\newcommand{\degree}{\ensuremath{^\circ}}	     % degree symbol
\newcommand{\kms}{ km\ s$^{-1}$}                     % km s-1
\usepackage{bm}
\usepackage{textcomp}
\newcommand{\Msun}{$M_{\odot}$}                       % M_o 
\newcommand{\Lsun}{$L_{\odot}$}                       % L_o 
\def\3he{$^3{\rm He}$}
\def\arcdeg{\hbox{$^\circ$}}
\def\lsim{\mathrel{\lower2.5pt\vbox{\lineskip=0pt\baselineskip=0pt
           \hbox{$<$}\hbox{$\sim$}}}}

\def\gsim{\mathrel{\lower2.5pt\vbox{\lineskip=0pt\baselineskip=0pt
           \hbox{$>$}\hbox{$\sim$}}}}

\def\spose#1{\hbox to 0pt{#1\hss}}
\def\simlt{\mathrel{\spose{\lower 3pt\hbox{$\mathchar"218$}}
     \raise 2.0pt\hbox{$\mathchar"13C$}}}
\def\simgt{\mathrel{\spose{\lower 3pt\hbox{$\mathchar"218$}}
     \raise 2.0pt\hbox{$\mathchar"13E$}}}

\begin{document}

\title{BLAST: The Mass Function, Lifetimes, and Properties of Intermediate Mass Cores from a 50 Square Degree Submillimeter Galactic Survey in Vela ($\ell \approx 265$\arcdeg)}
\shorttitle{COLD CORES IN VELA}
\shortauthors{NETTERFIELD ET EL.}

\author{
Calvin~B.~Netterfield\altaffilmark{1,2},
Peter~A.~R.~Ade\altaffilmark{3},
James~J.~Bock\altaffilmark{4}, 
Edward~L.~Chapin\altaffilmark{5}, 
Mark~J.~Devlin\altaffilmark{6},
Matthew~Griffin\altaffilmark{3},
Joshua~O.~Gundersen\altaffilmark{7}, 
Mark~Halpern\altaffilmark{5},
Peter~C.~Hargrave\altaffilmark{3},
David~H.~Hughes\altaffilmark{8}, 
Jeff~Klein\altaffilmark{6}, 
Gaelen~Marsden\altaffilmark{5}, 
Peter~G.~Martin\altaffilmark{1,9}, 
Phillip~Mauskopf\altaffilmark{3}, 
Luca~Olmi\altaffilmark{10,11},
Enzo~Pascale\altaffilmark{3}, 
Guillaume~Patanchon\altaffilmark{12}, 
Marie~Rex\altaffilmark{6}, 
Arabindo~Roy\altaffilmark{1}, 
Douglas~Scott\altaffilmark{5}, 
Christopher~Semisch\altaffilmark{6}, 
Nicholas~Thomas\altaffilmark{7}, 
Matthew~D.~P.~Truch\altaffilmark{6}, 
Carole~Tucker\altaffilmark{3}, 
Gregory~S.~Tucker\altaffilmark{13}, 
Marco~P.~Viero\altaffilmark{1},
Donald~V.~Wiebe\altaffilmark{2,5}
}

\altaffiltext{1}{Department of Astronomy \& Astrophysics, University of Toronto, 50
St. George Street Toronto, ON M5S~3H4, Canada}

\altaffiltext{2}{Department of Physics, University of Toronto, 60 St. George Street,
Toronto, ON M5S~1A7, Canada}

\altaffiltext{3}{School of Physics \& Astronomy, Cardiff University, 5 The Parade,
Cardiff, CF24 3AA, UK}  

\altaffiltext{4}{Jet Propulsion Laboratory, Pasadena, CA 91109-8099, USA}

\altaffiltext{5}{Department of Physics \& Astronomy, University of British Columbia,
6224 Agricultural Road, Vancouver, BC V6T 1Z1, Canada}

\altaffiltext{6}{Department of Physics \& Astronomy, University of Pennsylvania, 209
South 33rd Street, Philadelphia, PA, 19104, USA}

\altaffiltext{7}{Department of Physics, University of Miami, 1320 Campo Sano Drive,
Coral Gables, FL 33146, USA}

\altaffiltext{8}{Instituto Nacional de Astrof\'isica \'Optica y Electr\'onica (INAOE), 
Aptdo. Postal 51 y 72000 Puebla, Mexico}

\altaffiltext{9}{Canadian Institute for Theoretical Astrophysics, University of
Toronto, 60 St. George Street, Toronto, ON M5S~3H8, Canada}

\altaffiltext{10}{University of Puerto Rico, Rio Piedras Campus, Physics Dept., Box
23343, UPR station, Puerto Rico}

\altaffiltext{11}{IRA-INAF, Largo E. Fermi 5, I-50125 Firenze, Italy}

\altaffiltext{12}{Laboratoire APC, 10, rue Alice Domon et
L{\'e}onie Duquet 75205 Paris, France}

\altaffiltext{13}{Department of Physics, Brown University, 182 Hope Street, Providence,
RI 02912, USA}

\begin{abstract}
We present first results from an unbiased 50\,deg$^2$ submillimeter Galactic survey at 250, 350, and 500\,\micron\ from the 2006 flight of the Balloon-borne Large Aperture Submillimeter Telescope (BLAST).  The map has resolution ranging from 36\arcsec\  to 60\arcsec\ in the three submillimeter bands spanning the thermal emission peak of cold starless cores.  We determine the temperature, luminosity, and mass of more than one thousand compact sources in a range of evolutionary stages and an unbiased statistical characterization of the population.  From comparison with C$^{18}$O data, we find the dust opacity per gas mass, $\kappa r = $ 0.16\,cm$^{2}$g$^{-1}$ at 250\,\micron, for cold clumps.   We find that 2\% of the mass of the molecular gas over this diverse region 
is in cores colder than 14\,K, and that the mass function for these cold cores is consistent with a power law with index $\alpha = -3.22\pm0.14$ over the mass range $14\mbox{\,M}_\odot < M < 80\mbox{\,M}_\odot$.  Additionally, we infer a mass-dependent cold core lifetime of $t_c(M) = 4 \times 10^6 (M/20 \mbox{\,M}_\odot)^{-0.9}$\,years --- longer than what has been found in previous surveys of either low or high mass cores, and significantly longer than free fall or likely turbulent decay times.  This implies some form of non-thermal support for cold cores during this early stage of star formation.  

\end{abstract}


\section{Observations}

Dust comprises $\sim$\,1\% of the Interstellar Medium (ISM) by mass \citep{hildebrand1983}.  In denser regions, where star formation takes place, the dust absorbs essentially all of the Near Infra-Red (NIR) and visible radiation, and then re-radiates it at longer wavelengths.  Therefore, characterization of dust emission through submillimeter observations near the peak of the dust's thermal emission at 250 to 500\,\micron\ probes, in an unbiased way, the total energetics of such a system.  In the submillimeter the dust remains optically thin, so that even the centers of dense starless cores can be probed.  Observations in these critical 250 to 500\,\micron\ bands are very difficult, or impossible, from the ground, due to absorption by the Earth's atmosphere, even at the highest and driest terrestrial sites.  BLAST, which observes from a balloon at an altitude of 35\,km, above $\sim$\,99.5\,\% of the atmosphere, was developed to solve this problem.

\begin{sidewaysfigure*}[tbhp]
\includegraphics[width=\linewidth]{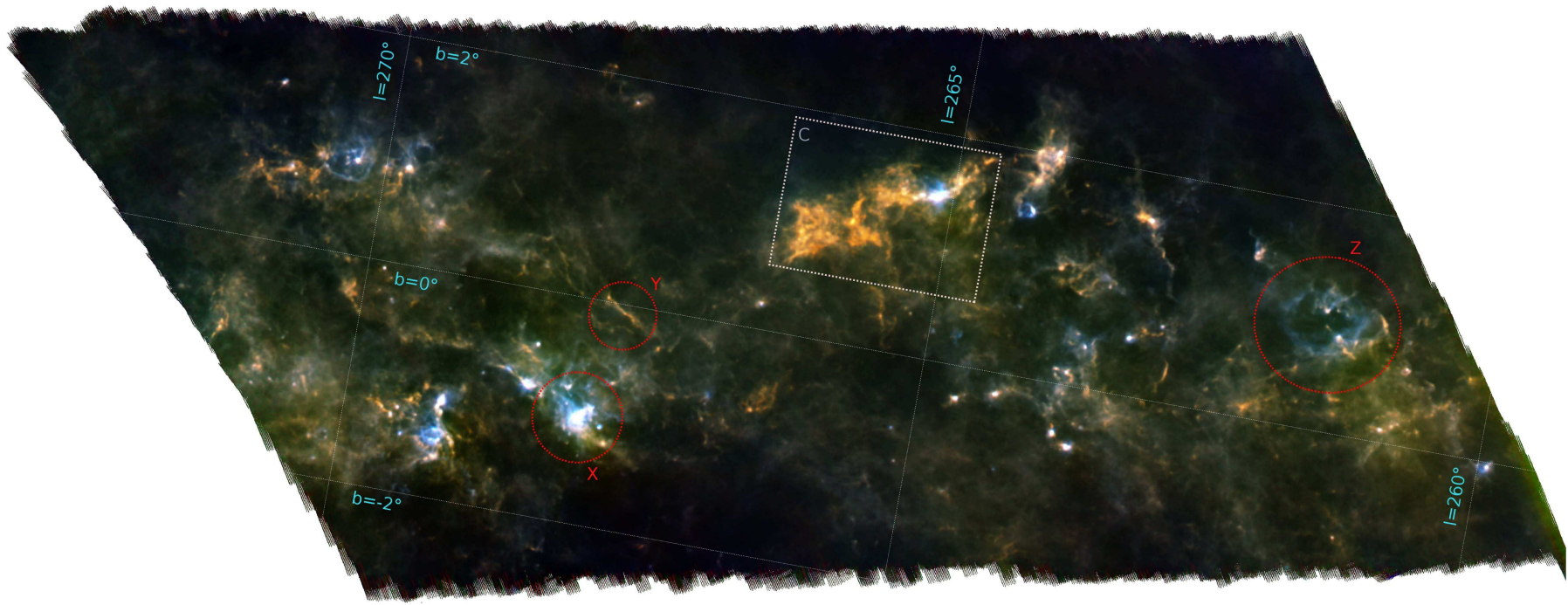}
\caption{False-color image of the full 50 square degree BLAST map of the Galactic plane towards Vela using the BLAST 250\,\micron\ channel for blue, the 350\,\micron\ channel for green, and the 500\,\micron\ channel for red.  Color in this image is an indicator of temperature, with blue regions being warmer ($\gsim$~25\,K) and red regions being cooler ($\lsim$~13\,K).  A non-linear color stretch has been employed to enhance the color contrast.
\label{fig:velafullcolor}}
\end{sidewaysfigure*}

In Figure~\ref{fig:velafullcolor}, we present a 50\,deg$^{2}$ submillimeter map of a portion of the Galactic plane made with BLAST\footnote[1]{The BLAST maps used in this paper are available for download from http://blastexperiment.info}. The map spans $\sim$\,10\degree\ in Galactic longitude and $\sim$\,5\degree\ in Galactic latitude in the constellation Vela, including the Vela Molecular Ridge (VMR). It is the result of 21\,hours of observing spaced throughout the 2006 flight of BLAST \citep{pascale2007}. 

This field, at Galactic longitude $\ell \approx260$ -- $270$\arcdeg, is dominated by relatively local emission, with the Perseus arm $\sim $10\,kpc behind it.  Most of the dust in this direction is in Giant Molecular Clouds (GMCs) in the VMR identified by CO emission \citep{murphy1991}.  Most of the region is thought to be 700\,$\pm$\,200\,pc away, though one of the clouds, centered at $\ell\approx 270$\arcdeg\ and $b\approx-1.5$\arcdeg\ is thought to be farther way ($\approx$ 2\,kpc), with a greater uncertainty in its distance \citep{liseau1992}.  For a variety of calculations, however, we rely on comparing BLAST extracted masses to those from \citet{yamaguchi1999}, who assume a distance of 700\,pc for the entire field.  For this reason we have assumed this same distance.

\begin{figure}[tbhp]
\includegraphics[width=\linewidth]{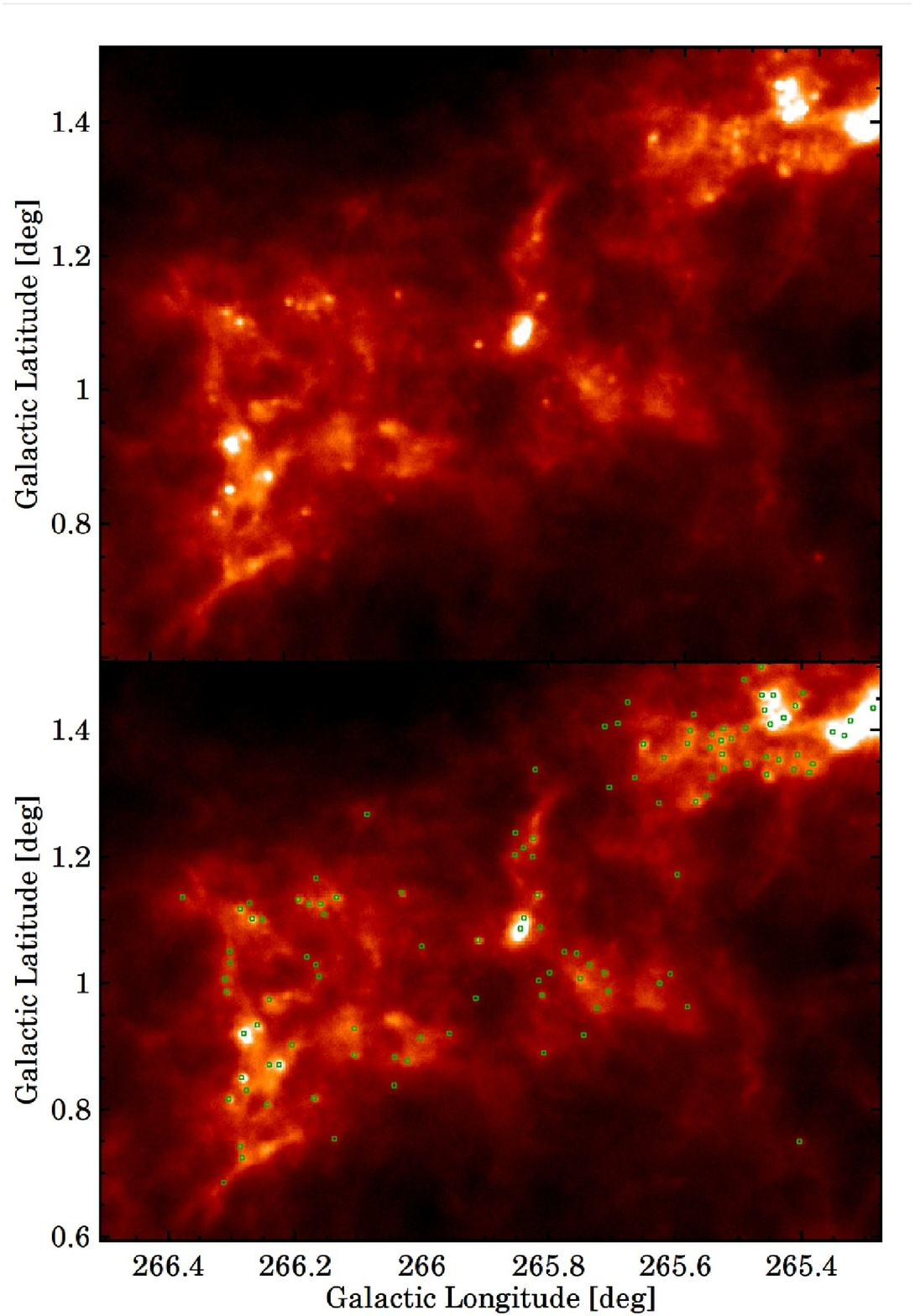}
\caption[Region C Closeup]
{A detail view of the 250\micron\ map of the left portion of region \emph{C} in Figure~\ref{fig:velafullcolor}.  This is a portion of Cloud 25 observed by \citet{yamaguchi1999} in C$^{18}$O, and found to be comprised of cold ($\sim10$\,K) gas with a mass of 38,000\,M$_\odot$.  Because of the low dust temperature, the cloud is essentially undetectable in the 100\,\micron\ \emph{IRAS} \citep{helou1988} maps.  This BLAST image, near the peak of the modified black-body spectrum, shows that the cloud is characterized by a large number of compact sources with an apparent characteristic size of 0.15\,pc.  The locations of the compact sources found by our algorithm are shown in the bottom panel.}
 \label{fig:velaC250u}
\end{figure}

The image contains objects at all stages of evolution, including: a large cool cloud with little evidence for recent star formation over much of it (region C in Figure~\ref{fig:velafullcolor}); cool dust arranged in linear structures (e.g., region Y in Figure~\ref{fig:velafullcolor} at $\ell$~=~267\fdg 7, $b=-$0\fdg 1); clouds which have been substantially heated by star formation (e.g., region X in Figure~\ref{fig:velafullcolor}, RCW~38, at $\ell$~=~267\fdg 9, $b=-$1\fdg 1); and the clumpy interstellar medium warmed up in a roughly circular structure by massive young stars (e.g., region Z in Figure~\ref{fig:velafullcolor}, containing RCW~32, at $\ell$~=~261\fdg 6, $b$~=~0\fdg 9).  Also very striking are the large numbers of compact sources throughout the image --- particularly in cooler regions (see Figure~\ref{fig:velaC250u}).  

The method by which a field is to be chosen can have significant impact on the interpretation of statistical results inferred from it.  For instance, if a field is chosen because of known on-going high mass star formation, then the amount of this star formation ongoing in the region will be biased.  These biases are reduced if the map is large, and eliminated if the nature of sources in the field is not considered in its selection.  Our selection of this field was not biased by its content; our goal was to make a 5\arcdeg\ by 10\arcdeg\ map of the Galactic plane (the size being set by our target depth).  The region which could be observed was highly constrained once the limited elevation range of the telescope, the location of the Sun given our flight time (shielding required that we stay more than 120\arcdeg\ in azimuth from the sun), and the range of latitude over which the payload could travel over its flight were considered.  We further required that the field be continuously observable so that multiple crossings of the map could be made at a variety of sidereal times, to provide cross-linking of the map.  These considerations limited us to the region we have chosen.

The map is also large, covering 10\arcdeg\ of Galactic longitude, or $\approx 3\%$ of the Galactic plane as projected on the sky.  Being large and not biased towards either active or inactive areas, the map probes the ISM over a wide range of physical conditions.  For these reasons, we consider it reasonable to take the molecular gas normalized core counts and mass function as an un-biased estimate of the mean properties of the Galaxy as a whole.

\section{The Instrument}
BLAST is described in detail in \citet{pascale2007} and the calibration in \citet{truch2009}, but it is summarized here.
BLAST has had two science flights: a four day flight from Kiruna, Sweden in June 2005 (BLAST05), and an twelve day flight from McMurdo Station, Antarctica in December 2006 (BLAST06).  The data presented here are from the 2006 flight.

BLAST06 is a 1.8-m Cassegrain telescope, whose under-illuminated primary
mirror produces beams with full width at half maximum  (FWHM) sizes of
36\arcsec, 42\arcsec, and 60\arcsec\ at 250, 350, and 500\,\micron,
respectively.  The camera consists of three
silicon-nitride ``spider web'' bolometer arrays \citep{rownd2003}
almost identical to those for SPIRE \citep{griffin2007} on {\it
Herschel}, with 149, 88, and 43 detectors at 250,
350, and 500\,\micron.  The 14\arcmin\ $\times$ 7\arcmin\ field-of-view
images the sky simultaneously in all three bands.
BLAST flies on a stratospheric balloon, at altitudes above 35\,km, to minimize in-band emission and absorption from the atmosphere.  

BLAST submillimeter observations are made by scanning the entire telescope in azimuth across the region to be mapped, with a slow drift in elevation.  For the observations described here, a speed of 0.2 degrees per second in azimuth was used.  The observations are repeated many times to fully sample the region.  Pointing reconstruction is provided by a pair of optical star cameras co-aligned with the submillimeter telescope which take images at the ends of the scans, when the telescope is nearly still.  Stars are identified within the optical image, and the position is reconstructed to better than 3\arcsec\ ($1\sigma$).  A set of fiber optic rate gyroscopes are integrated in between these star camera solutions.  The elevation-dependent relative orientation of the star cameras to the submillimeter telescope is determined from observations of known compact sources within the observing region.  The overall registration of the map is accurate to $<5\arcsec$.

\section{Data Analysis}
The raw BLAST data are reduced using a common pipeline.  The absolute gain of BLAST06, including antenna efficiency, was determined from regular observations of the evolved star VY~CMa.  Errors in the VY~CMa spectral energy distribution (SED) produce highly correlated absolute uncertainties of 10\%, 12\%, and 13\% at 250, 350, and 500\,\micron, respectively \citep{truch2009}.  Finally, maximum likelihood maps are made \citep{wiebe2008, patanchon2007}.  

\subsection{Source Extraction}
\label{sec:fits}

The source extraction and source property fitting techniques are similar to those used during analysis of the BLAST05 maps \citep{chapin2007}.  
A beam-equivalent-sized Mexican hat wavelet type convolution is applied to the 250 and 350\,\micron~maps, which identifies objects in the confused images by in effect subtracting a local background \citep[e.g., ][]{barnard2004}.  Peaks above a threshold in both maps form the candidate source list.  This technique finds 1549 candidate sources in the 250\,\micron\ band and 1302 candidate sources in the 350\,\micron\ band.  The 500\,\micron\ map is not used for source identification due to the greater source-source and source-background confusion resulting from the lower resolution.  

Circular Gaussians are fit to each candidate source in a 4\arcmin~diameter area extracted from the flux density map.  The center of the Gaussian fit is allowed to move at most by 20\arcsec\ relative to the candidate source location, and the FWHM to vary from 90\,\% of the beam size to 120\arcsec.  To eliminate contamination from the other sources, Gaussians are simultaneously fit to all candidate sources in the fit area.  A planar background is also fit.  Any poor fits (e.g., negative amplitudes, or parameters that reach the imposed limits) are discarded.  The 250 and 350\,\micron\ candidate lists, containing all locations, sizes and approximate flux densities, are merged.  Noise and background confusion may tend to shift the centroids of the fits.  Consequently, sources lying within 20\arcsec\ of each other are considered to be the same and the catalog records the position and size of the higher S/N object.  Sources appearing in only one of the two bands are also adopted in the final list.  In this procedure, 1109 sources were detected in the 250\,\micron\ band, and 920 sources were  detected in the 350\,\micron\ band.  There were 703 common to both bands, with a total of 1326 sources.

Gaussians are then refit using the fixed size (convolved to account for the differing beam sizes) and location parameters from the combined catalog, in all three wavelength maps.  Final fluxes are the integrals of these Gaussian fits.  Due to the sidelobes in the 250\,\micron\ beam there is a risk of the source finder identifying the sidelobes around bright sources. To prevent this, sources that are in the location and have the appropriate relative amplitude of sidelobes to a bright source are removed from the source list.  This removes 44 sources from the  250\,\micron\ list, leaving a total catalog of 1282 sources, as shown in Table~\ref{tab:measuredtable}.

\subsection{Simulations}
\label{sec:sims}

\begin{figure}[tbhp]
\includegraphics[angle=270,width=\linewidth]{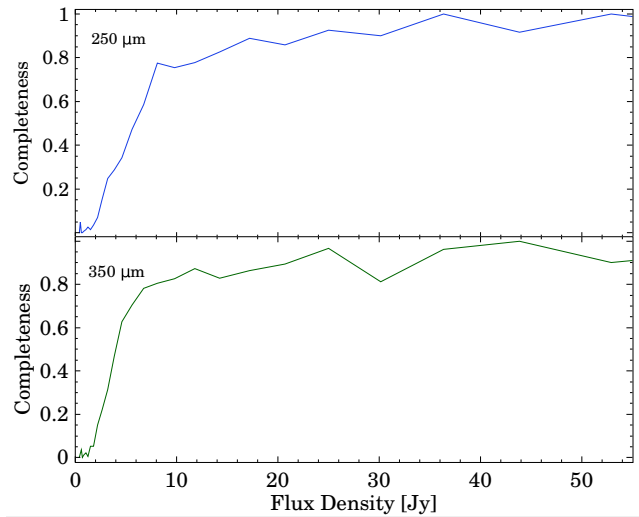}
\caption[completeness vs flux]
{Completeness as a function of flux density at 250\,\micron\ and 350\,\micron\ for simulated sources equal to or smaller than 60\arcsec, 
 the size of the median source found in the BLAST map.  As shown in Section~\ref{sec:sourceSizes}, completeness is lower for more 
diffuse sources, since at a given flux density level a larger source will have lower contrast.}
 \label{fig:completeness}
\end{figure}

Simulations are used to determine the bias and completeness of the source extraction routine.  New sources are added to the map by taking the final catalog, convolving them with the measured beam in each band and inserting them into the map.  Their locations are chosen to be at least 2\arcmin~from their original location, but not more than 5\arcmin, so that the fake sources will reside in a similar background environment to their original location.  These added sources are not allowed to overlap each other, but are not prevented from overlapping sources originally present in the map so that the simulation will account for errors due to confusion.  The resulting set of maps is run through the source extraction pipeline and the extracted source parameters are compared to the simulation input.  To probe completeness at the low flux and large size limit, additional input catalogs are generated with excess numbers of faint and extended sources.    The results of these simulations are shown in Figure~\ref{fig:completeness} and are later used to correct the size distribution in Figure~\ref{fig:sizes}.

The mean ratio of the recovered flux to the flux input to the simulation is determined to be 0.731, 0.911 and 1.120 at 250, 350, and 500\,\micron, respectively.
The largest source of this bias is that the telescope beams are not perfectly Gaussian.  Two sidelobes of the 250\,\micron\ beam contain $\sim$ 24\,\% of the power, while the flux extraction procedure only fits a Gaussian to the main lobe.  The probability of detection as a function of flux density is shown in Figure~\ref{fig:completeness}.   The completeness lines derived from this analysis are used to make cuts in mass later in the analysis.

As the artificial sources are placed into the map in the same environments as the real sources lie, we can test the bias of our estimated masses caused by confusion and source mixing.  We fit the SED (see Section~\ref{sec:sed}) to both the input and recovered source fluxes, and from this determine the mass (see Section~\ref{sec:mass}).  We find that the recovered masses are biased high by 12\%, which is small compared to the other uncertainties.   We conclude that confusion is not dominating our fluxes over the mass range of interest.

\subsection{Spectral Fits}
\label{sec:sed}
A single-temperature modified black-body SED is fit to the three BLAST fluxes for each source, from which we determine the temperature, mass, and luminosity.  The routine used here is identical to the one used for BLAST05 results \citep{chapin2007, truch2008}.  These fits require knowledge of the dust emissivity index, $\beta$.  The further determination of mass and luminosity requires knowledge of the dust mass absorption coefficient, $\kappa$, the dust to gas ratio, $r$, and the distance to the sources, $d$.  For the values we use for these parameters, and for details of the model, see Section~\ref{sec:sourceProperties}.

The uncertainties in temperature, mass, and luminosity, not including uncertainties in $\beta$, $\kappa$, $r$, and $d$, are obtained by performing Monte Carlo simulations.  Mock fluxes are generated for each source from Gaussian noise using the amplitude of the input source uncertainty and the known correlated and uncorrelated flux density calibration uncertainties.  An SED is fit to each of these mock sources.  The distribution of the fit parameters gives the 68\% confidence intervals.  An additional correction to the uncertainties is calculated based on the reduced $\chi^2$ for the simulated sources extracted from the map described above.  The uncertainties in temperature, luminosity, and mass are then scaled to compensate for any deviation from unity in the reduced $\chi^2$.  The results of these fits are presented in Table~\ref{tab:derivedtable}.

\section{Source Properties}
\label{sec:sourceProperties}

\subsection{Source Sizes}
\label{sec:sourceSizes}
\begin{figure}[tbhp]
\centering
\includegraphics[angle=270,width=\linewidth]{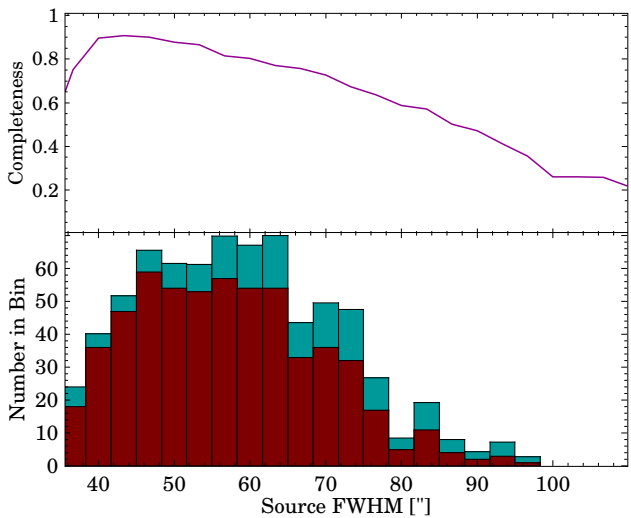}
\caption[Source sizes and size completeness]
{$\it{Top~panel:}$ Completeness as a function of 250\,\micron\ source FWHM, with source fluxes drawn from the actual distribution.  Completeness drops for higher source sizes.
 
$\it{Bottom~panel:}$ Source size histogram.  The lower (red) bars show the number of sources per bin, while the upper (blue) bins correct for the size completeness shown in the top panel.  Despite the decreasing completeness for more extended sources, there is no evidence for a large population of circular sources larger than 80\arcsec.}
 \label{fig:sizes}
\end{figure}

BLAST has a FWHM beam width of 36\arcsec\ at 250\,\micron.  The high signal-to-noise ratio of the map allows us to infer intrinsic source sizes below this scale by deconvolving the BLAST beam from the measured source FWHM.  We find that the sizes obtained from the fit (Figure~\ref{fig:sizes}) are broader than the intrinsic beam size, with a typical size of 62\arcsec, which corresponds to an intrinsic deconvolved source size of 0.15\,pc at the distance of the VMR, assuming Gaussian source profiles.  These sizes are recorded in Table~\ref{tab:derivedtable}.

\begin{figure}[tbhp]
\includegraphics[angle=270,width=\linewidth]{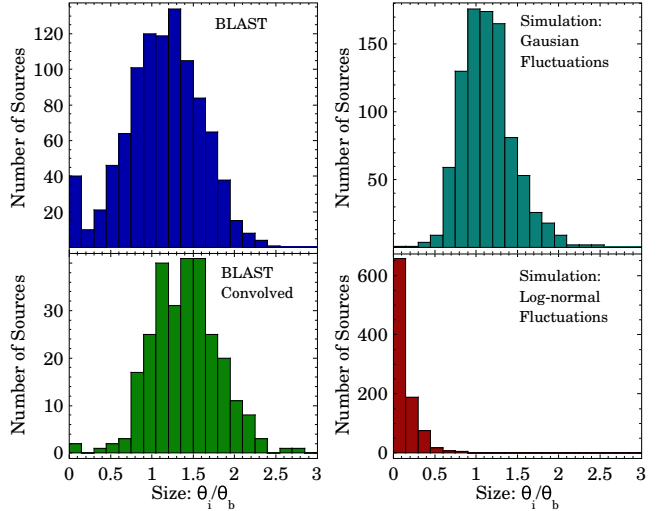}
\caption[Size Tests]
{Comparison of best fit source sizes.  $\theta_{\textrm{b}}$ is the BLAST beam size.  $\theta_{\textrm{i}} = (\theta_{\textrm{fit}}^2-\theta_{\textrm{b}}^2)^{1/2}$ is the inferred intrinsic source size.  \emph{Top Left:} BLAST 250\,\micron\ map.  \emph{Bottom Left:} BLAST map, convolved to 60\arcsec, and then re-binned into double-sized pixels, to simulate observing the same field at a greater distance. If we were truly resolving objects with well defined boundaries, one would expect the distribution in the bottom left panel to look like that in the top left, only shifted dramatically leftward.  Instead, the distribution looks similar.  \emph{Top Right:} The source fitter applied to a map of random Gaussian fluctuations which has been convolved by the BLAST beam.   \emph{Bottom Right:} The source fitter applied to a random log-normal fluctuations map which has been convolved by the BLAST beam. The size distribution of fit sources depends on the statistics of the map, as well as on the nature of the sources.}
 \label{fig:SizeTests}
\end{figure}

\begin{figure}[tbhp]
\includegraphics[angle=270,width=\linewidth]{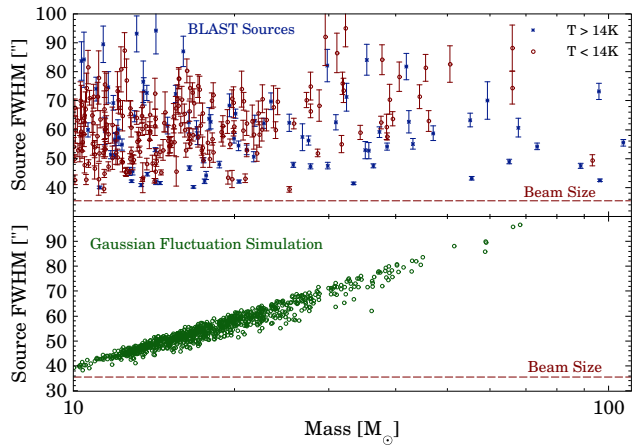}
\caption[Sources sizes as a function of mass]
{Source sizes in the 250um BLAST map as a function of mass.  The BLAST source sizes are independent of their mass, as would be expected if the size is a result of their power-law envelopes \citep{young2003}.  For comparison, the sizes from the Gaussian fluctuations simulated map grow as the square root of their mass, as would be expected from constant surface density.  From this result, it is clear that BLAST is not just fitting to Gaussian fluctuations; the results are consistent with fitting to the power law envelopes of cold cores.}
\label{fig:sizeVsMass}
\end{figure}

The size distribution of the sources is shown in Figure~\ref{fig:SizeTests}.  The distribution of source sizes from the BLAST map at 250\,\micron~is broad, with a peak at $\theta_{\textrm{i}}/\theta_{\textrm{b}} \approx 1.1$, where $\theta_{\textrm{b}}$ is the BLAST beam size and $\theta_{\textrm{i}} = (\theta_{\textrm{fit}}^2-\theta_{\textrm{b}}^2)^{1/2}$ is the inferred intrinsic source size.  The distribution falls off at larger scales faster than our completeness for rotationally symmetric Gaussian sources would imply (Figure~\ref{fig:sizes}).  From this result, it would be tempting to conclude that we have barely resolved the sources, which would have a typical size of $\approx0.15$~pc  at the assumed distance of $700$~pc.  However, since star-forming objects are not point sources, and are 
superimposed on structure covering a wide range of scales, one should be careful to test whether sources are really being resolved.

If we were barely resolving compact sources, then we would expect that if the region we were mapping had been further away, the sources would have appeared smaller, and the size distribution would have peaked at a smaller angular scale.  To test this, we have convolved 
the 250\,\micron~map to a resolution of 60\arcsec, and then re-binned it into double-sized pixels, to produce a map as we would have seen it, were the field approximately twice as far away.  As confusion becomes more of a problem at the lower resolution, we see fewer sources.  The size distribution of the sources we do see does not shift to smaller scales, and there are even fewer sources consistent with being point-like.  We would have expected that there would be many more.

Results where  $\theta_{\textrm{i}}/\theta_{\textrm{b}}$ is roughly independent of spatial resolution for sources in star forming regions has been seen elsewhere, and is a natural consequence of fitting Gaussians to beam-convolved cores which have intrinsic power-law envelopes.  Our size distribution is similar to what was seen in Perseus, Serpens, and Ophiuchus \citep{enoch2008}, and can be interpreted as being consistent with power-law envelopes with exponent $-2.0 < p < -1.0$ \citep{young2003}.

To test an entirely different possibility --- that apparent source sizes are intrinsic to fitting Gaussians to random Gaussian fluctuations --- we created two simulated maps, one with uncorrelated Gaussian fluctuations convolved with the beam, and one with uncorrelated log-normal fluctuations convolved with the beam.  The source sizes from the fits to the Gaussian fluctuation map have a qualitatively similar distribution to what is found in star forming regions, including what we have found here.  

To discriminate between these two possible interpretations, we note that if the resolution independent $\theta_{\textrm{i}}/\theta_{\textrm{b}}$ is due to fitting to power-law cores, then the fit size should be roughly independent of mass (as determined in Section~\ref{sec:mass}), while if it is due to fitting to random Gaussian fluctuations, we might expect a quadratic mass dependence on the size.  Figure~\ref{fig:sizeVsMass} shows that our sizes are in fact independent of mass.  For this reason, our results are consistent with fitting to cores whose envelopes drop off roughly as power-laws.   However, note that simple models of cold gravitationally bound but stable pre-stellar objects, such as Bonner-Ebert spheres, do not have power law envelopes.  Nor are power law envelopes the only explanation for the lack of a size-mass relationship, but this is beyond the scope of this paper.  

We conclude that \emph{size} and \emph{envelope} are poorly defined concepts for these sources.  They can not be regarded as isolated objects with well defined boundaries on a uniform background.   In determining the SED of such an object when the resolution varies between bands, it is important to insure that the flux is integrated assuming the same spatial size in each band (see Section \ref{sec:fits}).

\subsection{Temperatures}
Source temperatures are determined by fitting the BLAST SEDs to an optically thin, modified black-body:
\label{sec:T}
\begin{equation}
  S_{\nu} = A
  \left(\frac{\nu}{\nu_0}\right)^{\beta} B_{\nu}(T),
\label{eq:sed}
\end{equation}
where $A$ is the fit amplitude, $B_{\nu}(T)$ is the Planck function, $\beta$ is the dust emissivity index, and $c/\nu_0$~=~250\,\micron. 

Simple dust emission models predict that the dust emissivity index will approach $\beta = 2$ at long wavelengths.  More complex models, which include the properties of disordered materials \citep{Meny2007} provide a context within which $\beta$ can be less than $2$, but only at temperatures above $T\approx 20$\,K.

Observationally, $\beta\approx2$ is common, though the value of the effective dust emissivity index has been observed to be as low as $\beta \approx 1$ in higher density or warmer regions \citep{Meny2007}.  While it is very difficult to determine the intrinsic dust $\beta$\ without data significantly longward of the peak in the Rayleigh-Jeans regime, \citet{dupac2002} suggest a temperature/emissivity index inverse correlation, with low temperature cores having $\beta\approx2$.  

Observations of $\beta < 2$ for cold regions have been interpreted as being the result of temperature variations within the observed beam and along the line of sight.  A self-shielded core with possible star formation within it will almost certainly not be isothermal. Simulations \citep{Goncalves2004, Li2003} predict temperature gradients in cold cores of up to $10$\,K, which will lead to an effective best fit $\beta<2$, even if the intrinsic properties of the dust have the theoretically expected $\beta=2$.  In these cases, use of the effective best fit $\beta$ rather than the intrinsic $\beta$ will cause the mass to be significantly underestimated; we have found by simulations that when using the approximation of an isothermal fit, using the intrinsic $\beta$ rather than fitting for it minimizes the error in derived mass, even in the case of significant temperature variations. 

We chose to use the theoretically motivated value of $\beta=2$.  For reference, if we were to choose $\beta=1.5$, the best fit temperatures would increase by $\approx 10\%$. 

\begin{figure}[tbhp]
\includegraphics[angle=270,width=\linewidth]{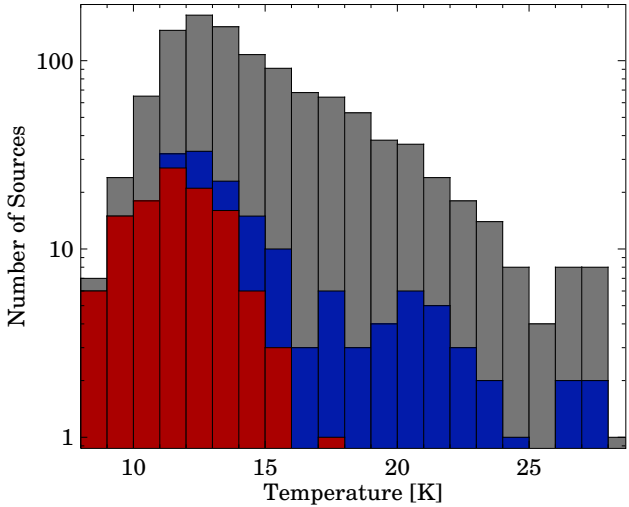}
\caption[Source temperature histogram]
{Histogram of source temperatures.  The upper (gray) bars represent the full catalog, the intermediate (blue) bars all sources with $M > 14$\,M$_\odot$, and the lower (red) bars all sources with $M>14$\,M$_\odot$\ and no \emph{IRAS} or \emph{MSX} associations (see Section~\ref{sec:context}).  Objects below 14\,K tend not to have IR counterparts and thus may be starless.  Typical uncertainties on an individual source temperature range from $\pm$ 0.75\,K for sources below 10\,K to $\pm ^{10}_{4}$\,K for sources above 25\,K.}
 \label{fig:T}
\end{figure}

BLAST is most effective at determining the temperatures of cooler ($\bm{<}$ 20\,K) sources, with typical uncertainties of 10\% for 11\,K sources, rising to over 40\% for 20\,K sources, as the peak of the modified black-body SED moves shortward of the BLAST bands.  This uncertainty  includes noise, confusion, and calibration uncertainty, but not systematic uncertainty from $\beta$.  The temperature distribution (Figure~\ref{fig:T}) has a strong peak at 12.5\,K, with a cutoff at the low end around 10\,K, and a tail extending to higher temperatures, with 19\% of the sources warmer than 20\,K.

\subsection{Source Mass}
\label{sec:mass}

Source mass is related to the amplitude fit in Equation~\ref{eq:sed} by
\begin{equation}
M_{\rm c} = \frac{Ad^2}{\kappa r}.
\label{eq:mass}
\end{equation}

The value of $\kappa r$ has not been well determined.  A summary of the values of $\kappa$ at 250\,\micron\ from the literature is shown in Table \ref{tab:kappas}.  The large spread of values follows the theoretically motivated variation in $\kappa$ between different regions and classes of objects.  We have determined $\kappa r$ for our region by comparing the BLAST dust emission to estimated gas mass from C$^{18}$O data.  This technique for empirically verifying $\kappa r$ by comparing submillimeter derived masses to those from CO has been applied previously by PRONAOS \citep{dupac2002}.  Although they do not calculate their own value of $\kappa r$, the PRONAOS team calculated masses based on both the \citet{draine2007} and \citet{ossenkopf1994} $\kappa r$ values, which bracket the mass estimates from external CO data for a few cold clouds.

\begin{deluxetable}{cc}
\tabletypesize{\scriptsize}
\tablecolumns{3}
\tablewidth{0pt}
\small
\tablecaption{Dust Opacity Values at 250\,\micron
\label{tab:kappas}
}
\tablehead{
 \colhead{$\kappa r$ (cm$^{2}$g$^{-1}$}) &
\colhead{Reference}
}
\startdata
0.024 & \citet{draine2007} \\
0.058 & \citet{desert1990} \\
0.1 & \citet{hildebrand1983} \\
 0.16 & This work \\
0.22--0.25 & \citet{ossenkopf1994} \\
\enddata
\tablecomments{Where the measurement is directly of $\kappa$, the canonical dust to gas ratio, $r = 0.01$, is assumed \citep{hildebrand1983} to provide a $\kappa r$ value.  The bottom entry was calculated for cold, dense regions, while the top is applicable to the diffuse ISM.}
 \end{deluxetable}

\begin{figure}[tbhp]
\includegraphics[width=\linewidth]{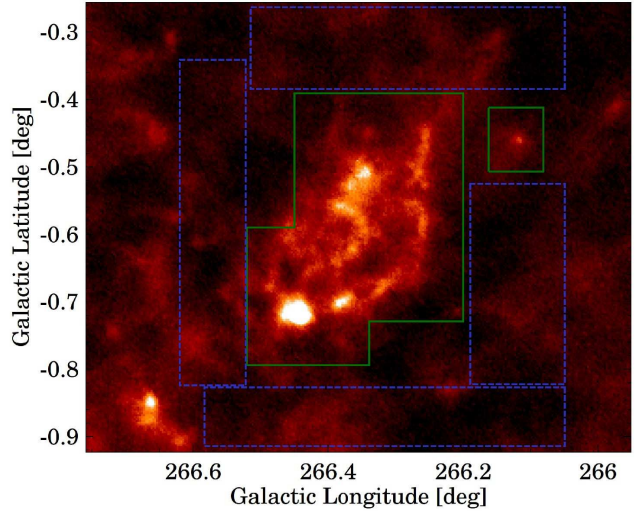}
\caption[cloud 28]
{Cloud 28, from \citet{yamaguchi1999} used to determine $\kappa r$.  The BLAST fluxes for this cloud were determined by summing the pixel values inside of the solid (green) region and subtracting an equal number of average baseline pixels determined from the dashed (blue) regions.}
 \label{fig:cloud28}
\end{figure}

The NANTEN instrument observed the $J =$ 1--0 $^{12}$CO and $^{13}$CO emission lines in the entire BLAST Vela field and the $J =$ 1--0 C$^{18}$O emission line for several targeted regions therein \citep{yamaguchi1999}.  One of these, the NANTEN cloud 28, is a well isolated, nearly isothermal 1400\,M$_\odot$ cold cloud, shown in the 250\,\micron\ BLAST band in Figure \ref{fig:cloud28}.    A single-temperature SED was fit to the BLAST-estimated fluxes and compared to the CO-derived mass to give us our BLAST-derived value of $\kappa r$.   In order to remain consistent with the NANTEN derived masses, we adopt a distance of $d=700$\,pc \citep{liseau1992}.  
 
We find $\kappa r = $ 0.16\,cm$^{2}$g$^{-1}$, or $\kappa =$ 16\,cm$^{2}$g$^{-1}$ assuming $r = 0.01$.  The procedure was repeated to see how much the compact source affects the result, by first removing the bright point source in Figure~\ref{fig:cloud28}, fitting its mass separately, and adding it back into the total.  This had a negligible ($\lsim$ 2\,\%) effect on our value for $\kappa r$.  

As a consistency check, we repeated this analysis with the largest cloud in \citet{yamaguchi1999} --- cloud 25, region C in Figure~\ref{fig:velafullcolor} --- though it is not as isothermal as cloud 28.  We separately fit the entire hot central region (centered on RCW 36) and the rest of the cloud.  When we do this we derive a temperature of 20.2\,K for the central region and 11.5\,K for the rest of the cloud and a $\kappa r$ value of $0.14$\,cm$^{2}$g$^{-1}$, quite consistent with the results from cloud 28, which we use in our analysis.

The value of $\kappa r$ is independent of the distance to the cloud, as both the C$^{18}$O and BLAST dust measurements have the same dependence on distance.  However, the value is strongly dependent on our assumed $\beta=2$.  Changing our assumption to $\beta=1.5$ changes $\kappa r$ by about 50\%.  However, if the dust properties are the same for our sources as for the clouds we have used for this calibration of $\kappa r$, then this will not affect their final masses.  Our biggest source of uncertainty is from the CO mass estimates --- these uncertainties are typically a factor of 1.5 to 2, corresponding to an uncertainty in $\kappa$ and our derived source masses by the same factor.

\begin{figure}[tbhp]
\includegraphics[angle=270,width=\linewidth]{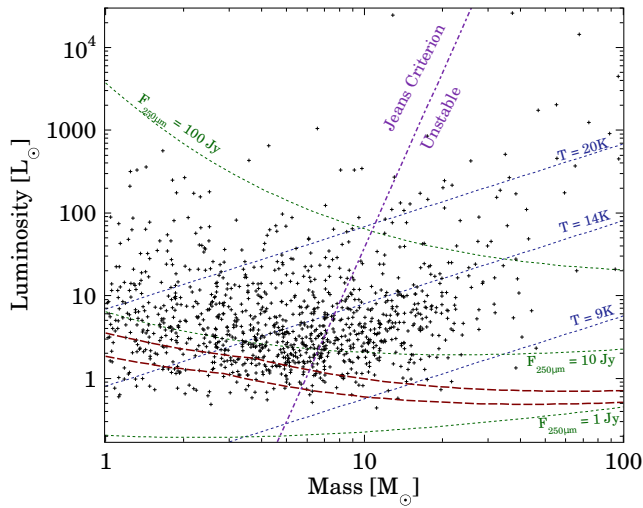}
\caption{Masses and luminosities of BLAST sources.  Individual BLAST sources are plotted in black.  Relative to the limiting completeness at high flux density, the catalog is $\bm{>}$~50\,\% complete above the lower dashed red line and $\bm{>}$~80\,\% complete above the upper dashed red line.   The Jeans stability criterion for 0.15 pc sources is shown by the diagonal dashed magenta line: sources to the right of the line exceed their Jeans mass and are unstable without some form of non-thermal support.  As discussed in Section~\ref{sec:context}, sources with $M>14$\,M$_\odot$ and $T<14$\,K do not typically have MSX counterparts - consistent with their being starless.  Their cold temperatures is consistent with a lack of a significant internal source of energy.
 \label{fig:lm}}
\end{figure}

The masses, luminosities, and temperatures of the cores are presented in Figure~\ref{fig:lm}.  The sources span the mass range from 1 to 100\,M$_\odot$, with most of the sources centered around their Jeans mass (see Section~\ref{sec:timescales}), which is several solar masses.  Due to confusion with background structure, the survey is incomplete for cooler sources at low mass.  Above 14\,M$_\odot$, it is complete for even the coolest sources.

In the same way that the power-law envelopes of these cores make size a poorly defined concept, the mass that we measure is a function of our spatial resolution.  This needs to be taken into consideration when comparing our masses to those measured in other fields.  Under the model that we are fitting to beam-convolved  $p \approx -1.6$ power law envelopes, simulations show that if we had mapped the same sources with the same (improved) spatial resolution as was done by Bolocam in Serpens \citep{enoch2008}, we would have found $\approx$ 1/2 of the mass that we found with our spatial resolution in Vela, because the fit would not have picked up as much of the envelopes.

\begin{figure}[tbhp]
\includegraphics[angle=270,width=\linewidth]{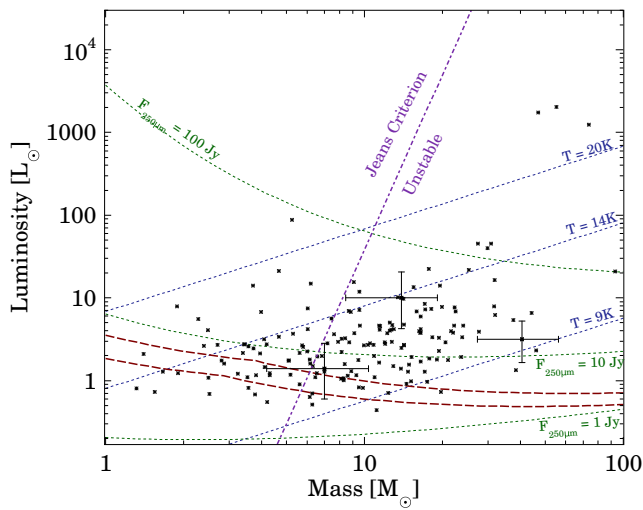}
\caption{Like Figure~\ref{fig:lm} but restricted to region C in Figure~\ref{fig:velafullcolor}.
 In addition, 1-$\sigma$ mass and luminosity error bars for three representative BLAST sources are plotted for reference.
\label{fig:lmVelaC}}
\end{figure}

In Figure~\ref{fig:lmVelaC}, we present the same information for sources within cloud 25 (object C in Figure~\ref{fig:velafullcolor}) which is a large cold cloud not visible in \emph{IRAS} \citep{helou1988} at 100\,\micron.  This is the largest of a class of objects visible in the map which are characterized by large numbers of cold sources with few mid-IR or far-IR counterparts.  

\section{Placing the Sources in Context}
\label{sec:context}
The sources found here bear a strong resemblance to the sources which have been identified elsewhere as starless cores. For example, they have a similar mass function, size, and temperature to cores in the Pipe Nebula \citep{alves2007, rathborne2008}, Perseus \citep{enoch2006}, Ophiuchus \citep{young2006}, and Serpens \citep{enoch2007}.   In these fields, which are in nearby clouds, sources have been found to have characteristic sizes between 0.05\,pc and 0.15\,pc.  As we have noted, in the context of power-law envelopes, these sizes are related to the spatial resolution of the instrument, and the steepness of the power-law, rather than to the size of the source; our derivation of larger spatial sizes from BLAST observations could be a consequence of fitting further down the power law envelope.  A high source density could cause source blending, but our simulations show that at the current detection threshold, this does not dominate.  For these reasons, we conclude that our objects are a mix of protostellar and starless cores, as has been seen in previous papers, but extending to higher masses due to our larger coverage volume.  Our sources are in the size range to be classified as cores rather than clumps according to the definitions used by \citet{motte2007} in looking at high mass star formation in Cygnus X.

It is not possible for us to discriminate unequivocally between protostellar and starless cores, but it is possible to use temperature as a rough discriminator.  Working under the assumption of single-core collapse in massive star formation, the class I protostellar core candidates in \citet{molinari2008} are all less than 20\,K.   Modeled dust temperatures for Class\,0 cores are $\approx$15\,K \citep{shirley2002}, and fully starless cores have been assumed to be at $\approx$10\,K \citep{enoch2008}.  For reference, low $A_v$ dust in the local Galactic neighborhood comes to an equilibrium temperature of $\approx$18\,K \citep{schlegel1998}.  Being externally heated, starless cores are colder than this due both being embedded in molecular clouds, and their self shielding.  

As a test of this rough classification, we can compare our catalog with other catalogs at shorter wavelengths.  \citet{baba2006} present a catalog of protostellar candidates in the GMC indicated as object C in Figure~\ref{fig:velafullcolor}, using \emph{IRAS}, \emph{MSX} \citep{egan2003}, and NIR data.   They find 30 candidate protostars, which, from their \emph{IRAS}, \emph{MSX}, and NIR fluxes, they interpret as being near the transition between Class I and Class II objects, with bolometric luminosities between 5 and 105 L$_\odot$ and protostar masses between 1 and 4 M$_\odot$.  Their stated completeness range is $L>5$\,L$_\odot$.

Of the 169 BLAST cores in object C, only 30 have any association with either an \emph{IRAS}  or \emph{MSX}  source, and of these, only 20 (12\%) are associated with the \citet{baba2006} protostar candidates.  The mean temperature of the protostar candidates is 14\,K,  $\approx 2$\,K warmer than cores without protostellar associations.  Taking the \citet{baba2004} interpretation of these objects at face value suggests that the transition from the intermediate mass analogs of Class I to Class II protostars is at around 14\,K, considerably lower than what is reported in \citet{molinari2008}, so the precise interpretation of these objects remains uncertain.  For this reason, while we expect that cooler cores will, on average, be at an earlier stage of star formation than warmer cores, we do not attempt to assign a protostellar class to our cores on the basis of temperature.

We extend our analysis of the association of BLAST sources to the \emph{IRAS} and \emph{MSX} catalogs to the entire map.  We consider a BLAST source to have a mid-IR or far-IR counterpart if a source in the \emph{IRAS} Point Source Catalog (Version 2.0) lies within $1\arcmin$ of a BLAST core, or if a source in the \emph{MSX} point source catalog (Version 2.3) lies within the BLAST source radius.  We find, in Figure~\ref{fig:T},  that for $M>14$\,M$_\odot$, the temperature histogram of sources warmer than 14\,K is dominated by sources with mid-IR or far-IR counterparts, but sources cooler than 14\,K are dominated by sources without them.  From this we define \emph{cold cores} as being cores with a dust temperature below 14\,K.  Their low temperatures and the lack of a mid-IR or far-IR counterpart indicates that they are likely to represent the earlier stages of star formation.  It is possible that low mass stars could be embedded within them without being detected, but the main point is that these cold cores do not appear to have a major internal energy source from accretion or nuclear burning.

\section{The Core Mass Function}
\label{sec:massfunction}
\begin{figure}[tbhp]
\includegraphics[angle=270,width=\linewidth]{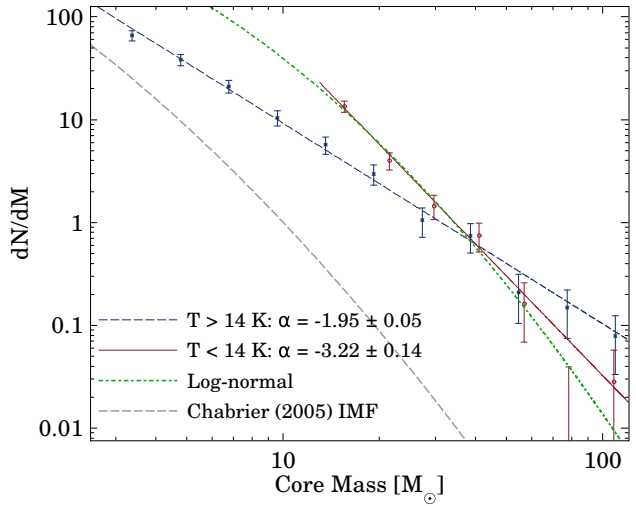}
\caption{The mass distribution of sources.  For sources warmer than 8.5\,K, the catalog is complete above 14\,M$_\odot$, and for sources warmer than 14\,K, the catalog is complete above 2.5\,M$_\odot$.  Data outside these ranges are not shown.
The masses in this plot are calculated assuming a dust emissivity index, $\beta=$\,2.0, a dust mass absorption coefficient, $\kappa r=0.16\,\mbox{cm}^2\mbox{g}^{-1}$, and a distance of 700\,pc.  Poisson uncertainties are shown, which do not include the uncertainty in these coefficients. A log normal distribution with $M_0=2.0$\,M$_\odot$\ and $\sigma=0.46$, normalized to fit our data, and the nearly parallel  \citet{chabrier2005} stellar IMF with arbitrary normalization are shown for reference.
\label{fig:hist_M}}
\end{figure}

The mass function of our cores is shown in Figure~\ref{fig:hist_M}.  Cores above 14\,M$_\odot$, for which our data are complete, follow a core mass function (CMF) of the form $dN/dM = n_0 M^{\alpha}$ with $\alpha = -2.77 \pm 0.16$.  However, this is a very heterogeneous collection of objects, at a wide range of evolutionary states.  

If, instead, we restrict our analysis to cold cores,  below 14\,K, we find a steeper index of $\alpha = -3.22 \pm 0.14$ and $n_0 = 9.0 \times 10^4$.  This index is steeper than has been seen in other fields at smaller scales; \citet{enoch2008} find $\alpha=-2.3\pm0.4$ over the mass interval $0.8 < M/\mbox{M}_\odot < 6$.  This steepening with increased mass would be expected if the mass function is well described by a log-normal distribution.

In contrast, if we restrict our analysis to sources which are above 14\,K, we see a significantly shallower index of $\alpha = -1.95\pm0.05$.  For these warmer sources our data are complete down to 2.5\,M$_\odot$, allowing the mass function to be evaluated over a wider range.  That warmer, more evolved cores follow a shallower mass function than cold cores has also been seen by \citet{enoch2008}, and is equivalent to the observation that high mass cores apparently spend a fractionally smaller amount of time in the cold phase than intermediate mass cores do. 

The distances to individual sources are uncertain, but if all of the sources we see are themselves drawn from a power-law distribution with the same index, the index of the mass spectrum will not be changed by errors in the distance estimates.  This will be true, for example, if some of the sources are further away, but still in the mass range described by the $\alpha = -3.2$ power law.  So, relative errors in our mass estimates  of only a factor of a few will not change the power law index, but if a substantial fraction of our sources are more distant and of a different type, then the interpretation could be compromised. 

\begin{figure}[tbhp]
\includegraphics[angle=270,width=\linewidth]{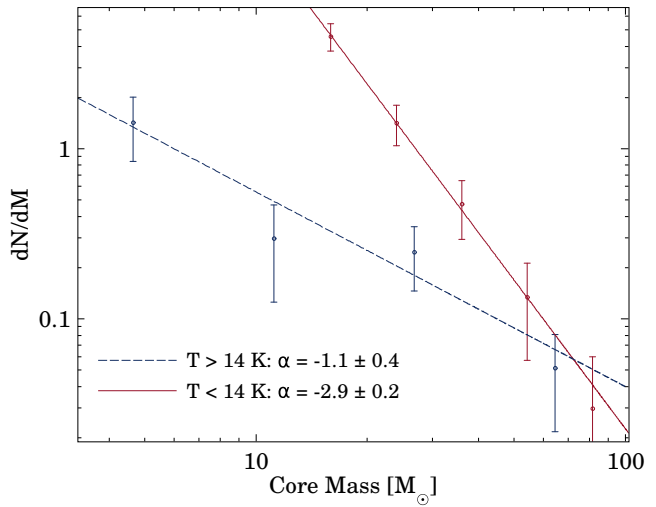}
\caption{The mass distribution of sources in region C, prepared with the same assumptions as Figure~\ref{fig:hist_M}.
\label{fig:hist_MC}}
\end{figure}

In order to minimize the possibility of strongly heterogeneous distances, in Figure~\ref{fig:hist_MC} we repeat the mass function analysis for sources in region C, which appears to be a single cloud with minimal scatter in distance.  We find the same trend, where the cold cores have a steeper mass function index than the warmer sources.  The cold core mass function for this single cold cloud is consistent, within the errors, with our results for the map as a whole.  This consistency between the cold CMF in the single cloud, and the cold CMF over the entire
region, which includes areas in a wide range of evolutionary states implies that cold CMF does not evolve significantly, when only cold cores
are considered.

We note that our observations of the cold cores are also consistent with predictions that the CMF should be well described by a log-normal distribution \citep{goodwin2008} of the form
\begin{equation}
\frac{dN}{dM} = \frac{A}{M}\exp\left(-\frac{(\log_{10}(M/M_0)^2}{2\sigma^2}\right).
\label{eq:lognormal}
\end{equation}
 With BLAST data alone, which are not complete down to the peak of a log-normal CMF, there is a strong degeneracy between $M_0$ and $\sigma$.   Taking $M_0 = 1$\,M$_\odot$\ from \citet{enoch2008} but correcting the masses by a factor of 2 for the different spatial resolution between our field and theirs, we find $\sigma\approx0.46$ for the whole BLAST field --- consistent with the their estimate of $\sigma = 0.31\pm0.23$.  Note also that \citet{enoch2008}, having data at a single frequency, do not measure temperature, which increases the errors in their masses.

\citet{alves2007} found that the mass function of cores at lower masses in the Pipe Nebula follows that of the stellar initial mass function, but at a factor of three higher mass.  From this, they inferred a star formation efficiency of $\epsilon=0.3$, under the interpretation that each core forms a single star.  While the stellar IMF for intermediate and high mass stars is fairly uncertain, our results remain consistent with this claim; the power law fit to the mass function for stars more massive than $1 \mbox{M}_\odot$ summarized in \citet{kroupa2007} gives $\alpha = -2.7 \pm 0.7$ if binary-companions are corrected for (or $\alpha = -2.3 \pm 0.7$ if they are not), which is statistically consistent with our measurement of $\alpha = -3.22\pm0.14$.  Similarly, the application of a shift in mass of a factor of four to the the log-normal fit to the IMF of \citet{chabrier2005} as proposed by \citet{alves2007}, is consistent with our data.

The observation that the CMF resembles the mass scaled stellar IMF has been interpreted as revealing the source of the IMF,  under the assumption that cores become stars with mass-independent multiplicity and efficiency.  This interpretation is equivalent to the assumption that the characteristic time for cold cores to become stars is independent of mass.  If the characteristic time for cold cores to become stars is not mass-independent, then the cold CMF must evolve (and we have not observed it in its time averaged steady state, despite the large area and wide range of evolutionary states represented by our map), or efficiency and multiplicity are functions of mass --- implying that any observed similarity is a coincidence.

\subsection{Fraction of Mass in Cold Cores}
Comparing the mass we find in cold cores with the molecular gas mass found in the same region in $^{12}$CO, we can determine the fraction of mass in cold cores.  As distance uncertainty affects both the core mass estimates and the gas mass estimates in the same way, this ratio is independent of our assumed distance, as long as the cores are well associated with structures in CO (visually, they are) and as long as our sample is not contaminated by distant objects of another type posing as cold cores.  The fact that the mass function falls so steeply (faster than $d^2$) means that the density of high mass distant objects is spatially much lower than the density of low mass closer objects --- reducing the probability that this is a dominant effect.  Our catalog contains 6,000\,M$_\odot$ of cores with $T<14$\,K.  To estimate the total mass in our cold cores, compensating for completeness, we integrate the log-normal mass function from Figure~\ref{fig:hist_M} to find a total mass in cores of 12,000\,M$_\odot$, compared with a total gas mass in CO in the VMR \citep{yamaguchi1999} of $5.6\times 10^5$\,M$_\odot$.  The total gas fraction in cold cores is thus estimated at 2\%.  Given the large and heterogeneous nature of the field, we suggest that this fraction can be taken as representative of the Galaxy as a whole.

\section{Lifetimes}
\label{sec:timescales}
\subsection{Gravitational Collapse}
Given the masses, sizes, and temperatures of the cores, we can calculate whether they can form pressure-supported structures which are stable against the Jeans instability \citep{Jeans1902}.  The maximum mass, $M$, enclosed within a radius, $R$, of such a structure is given by 
\begin{equation}
\left(M/R\right)_\textrm{J} = 4.2 \frac{c_\textrm{s}^2}{G}
\label{eq:mbe}
\end{equation}
where $c_s = \sqrt{kT/2.3 m_\textrm{p}}$ is the speed of sound in the medium, $k$ is Boltzman's constant, $T$ is the gas temperature, which we assume to be equal to the dust temperature, and $m_p$ is the proton mass.  For a Gaussian source, the enclosed $M/R$ has a peak at $2.135\sigma$ where the enclosed mass is 79.3\% of the total fit mass.  A truncated power law core has a larger $M/R$ than a Gaussian, with a maximum which is dependent on the size of the inner knee.  Since we are fitting Gaussians, and do not have the resolution to determine the inner knee radius, we take the conservative position that the maximum $M/R$ for our cores is greater than that of a Gaussian.  The criterion for Gaussian cores is plotted in Figure~\ref{fig:lm} for our median sized (0.15\,pc) sources.

Restricting ourselves to sources with temperatures that are consistent with an early stage of star formation ($T<14$\,K), and considering the size of each source, we find a median $M/M_J = 1.8$, meaning that the distribution peaks near the Jeans criterion, but that most of the sources (80\%) are unstable.  If we further restrict ourselves to the mass range for which the survey is complete ($M>14$\,M$_\odot$), our sources exceed their Jeans masses by a median factor of $M/M_J = 4$.  This ratio is linear in the assumed distance, so even if the cloud were at 500\,pc, rather than 700\,pc, the majority of the sources would still be unstable.  Sources more distant than the assumed 700\,pc exceed their Jeans mass by an even larger factor.  

Without some other form of support, cores that exceed the Jeans mass would collapse to form stars in a free-fall time, given by \citep{mckee2007},
\begin{equation}
t_{\rm{ff}} = 1.37 \times 10^{6}\rm{\,yr}\left[\frac{10^3 \rm{\,cm}^{-3}}{\it{\bar{n}}}\right]^{1/2}.
\label{eq:tff}
\end{equation}
For the $M > 14$\,M$_\odot$, $T < 14$\,K intermediate mass cores in our catalog, the median density is $\bar{n}=1.7\times10^4$\,cm$^{-3}$.  The free-fall times are roughly Gaussian distributed as $t_{\rm{ff}} = (3.4 \pm 1.2) \times 10^5$\,years.  This is a lower limit to the lifetimes of cold cores, in that it is possible that many of them may contain pre-Class II protostars.

These characteristic times do not fit comfortably with the character of the GMCs within which the sources lie.  Within region C in Figure~\ref{fig:velafullcolor}, one large cluster
containing more than 350 stars has formed two to three million years ago \citep{baba2004}, which implies that the GMC is older than this.  Given this age, the free-fall times, and the large number of cold cores in region C, one would expect that the GMC would contain stars at various stages of evolution.  However, besides two very small candidate star clusters, and $\sim$ 30 candidate protostars identified in the mid-IR by MSX and confirmed in the NIR \citep{baba2006}, there is little sign of ongoing star formation.  Only 6\% of the cores with mass over 14\,M$_\odot$ in this cloud have mid-IR (\emph{MSX}) counterparts, compared to 22\% considering the entire map.  Since \citet{baba2006} argue that \emph{MSX} has the sensitivity to find the intermediate mass protostars which our $M>14$\,M$_\odot$ will become, the cloud appears to be in a very early stage of evolution.  This suggests that we are either seeing the cloud at a very special time, or that the lifetimes of the cores are much longer than the free fall times, and more on the order of the few million year minimum age of the GMC.

\subsection{Gas Consumption}
The characteristic time for cold cores to become stars can also be estimated from the fraction of the molecular gas in cold cores, by assuming that all cold cores eventually become stars with a mass-independent efficiency, and from the molecular gas depletion time, $t_{\textrm{H}_2} = M/\textrm{SFR}$.  

A study of star formation in several nearby galaxies has found that H$_2$ in spirals forms stars at constant efficiency over a range of scales and physical conditions, with a molecular gas depletion time of $t_{\textrm{H}_2} = (2 \pm 0.8) \times 10^{9}$ years \citep{bigiel2008}.  The galaxy averaged molecular gas depletion time for the Milky Way, including the consumption of helium, can be estimated from the total molecular gas mass $M_{\textrm{H}_2}\approx1.3\times 10^9$\,M$_\odot$ and star formation rate $R_\textrm{MW} \approx 2.7$\,M$_\odot$/yr \citep{misiriotis2006} to be $t_{\textrm{H}_2} = M_{\textrm{H}_2}/0.73R_{\textrm{MW}} \approx 6.6\times10^8$ years.  This estimate for the Milky Way gas depletion time is substantially less than for the nearby galaxies, but given the considerable uncertainties, roughly consistent.  For the analysis that follows, we will use the Milky Way gas depletion time, but note that using $t_{\textrm{H}_2}$ from the nearby galaxies would give a substantially larger characteristic times.

If we further extrapolate from our observations over a broad region that 2\% of molecular gas in the entire Milky Way is in cold cores, and if we assume that these cores will eventually form stars with an efficiency $\epsilon=0.3$ \citep{matzner2000}, the mass-independent lifetime for cold cores is given by

\begin{equation}
t_c =  0.02\epsilon\,t_{\textrm{H}_2} = 4\times10^6\,\mbox{yr.}
\label{eqn:tau1}
\end{equation}

If a large fraction of cold cores never becomes stars (effectively reducing the efficiency), then this lifetime would be an over-estimate of those that do produce a star.

\subsection{Mass Dependence}
We can further develop a toy model to determine the mass-dependent lifetime for the cores using the core mass function.
We first estimate the expected star formation rate per mass interval for our region by assuming that star formation follows gas mass, reversing the steps in the previous section.  Measurements of $^{12}$CO give a total molecular gas mass in the VMR \citep{yamaguchi1999} of $5.6\times 10^5$\,M$_\odot$, compared to a total molecular gas mass in the Galaxy of $1.3\times 10^9$\,M$_\odot$.  Since star formation follows molecular gas, our map accounts for 0.043\% of the total star formation of the Galaxy.  So, given a Galactic star formation rate of 2.7\,M$_\odot$ per year, this region of the Galaxy produces stars at a rate of $1.2\times10^{-3}$\,M$_\odot\rm{yr}^{-1}$.  With a canonical initial stellar mass function for stars, neglecting multiplicity \citep{kroupa2007}, normalized by the total star formation rate in Vela, the star formation rate per mass interval per year in Vela is
\begin{equation}
F_{{\rm vela}} = \frac{dN_s}{dM_sdt} = 2.6\times 10^{-4} M_s^{-2.3} \mbox{\,[M}_\odot^{-1} \mbox{yr}^{-1}\mbox{]}
\label{eqn:rv}
\end{equation}
over the mass interval $0.5 \mbox{\,M}_\odot < M < 150 \mbox{\,M}_\odot$.  

Assuming a power law core mass function, $dN_c/dM_c = n_0M_c^\alpha$, and that each core will form $k$  stars of mass $M_s$ with efficiency $\epsilon$, the mass function of stars which will eventually be created from the cores which currently exist is

\begin{align}
\left(\frac{dN_s}{dM_s}\right)_P &= \left(\frac{dN_c}{dM_c}\right)\left(\frac{dN_s}{dN_c}\right)\left(\frac{dM_c}{dM_s}\right) \\
             &= \left(n_0M_c^\alpha\right)\left(k\right)\left(\frac{k}{\epsilon}\right) \\
             &= \frac{n_0k^{\alpha+2}}{\epsilon^{\alpha + 1}}M_s^\alpha
\label{eqn:psmf}
\end{align}. 

Further assuming a steady state for both the core mass function and the normalized star formation rate, $R_\textrm{vela}$, when averaged over the whole region, the characteristic time for the cores to form stars is

\begin{align}
t_c(M) &= \frac{(dN_s/dM_s)_P}{F_\textrm{vela}} \\
  &= \frac{n_0\epsilon^{1.3}}{2.6\times 10^{-4} k^{0.3}} M_{{\rm core}}^{\alpha + 2.3}.
\end{align}
In Section~\ref{sec:massfunction} we showed that over the range $14<M/\mbox{M}_\odot<100$, $n_0 = 9.0\times10^4$ and $\alpha = -3.22$.
Assuming $k=2$ and $\epsilon = 0.3$ we find
\begin{align}
t_c(M) = 4 \times 10^6 (M/20\mbox{\,M}_\odot)^{-0.9}\mbox{\,yr.}
\label{eqn:tau2}
\end{align}
The core mass function being steeper than the Salpeter mass function for stars implies that high mass cores evolve more quickly than low mass cores.  This time dependence is reliant on the highly uncertain IMF spectral index.  A steeper IMF would imply less time dependence.

These times are sensitive to uncertainties in the $^{12}$CO masses used to estimate the fraction of gas in molecular clouds, and the molecular gas depletion time for Vela.  Using $t_{\textrm{H}_2} = (2\pm0.8)\times10^9$\,years from \citet{bigiel2008} instead of the estimate from the very uncertain values from the Milky Way gives $t_c = (12\pm5)\times10^6$\,years.  These uncertainties could substantially change the estimate, but nevertheless, it would be very difficult to reconcile our counts with free-fall times.  

The few million year lifetime is roughly consistent with a naive crossing time based on 0.1\,\kms\ infall velocities measured in low mass cold cores \citep{bergin2007}.  

The lifetimes of these intermediate cold cores are longer but roughly consistent with previous estimates for low mass cores.  A compilation of previous data reported in \citet{WardThompson2007} finds lifetimes for densities like we find of around a million years, which is several times the free-fall time.  Similarly, lifetime estimates of low mass starless and protostellar cores in Perseus, Serpens, and Ophiuchus based on the ratio of starless, Class 0 and Class I cores to Class II cores gives starless core lifetimes of a half million years, and combined starless and Class 0/I protostellar lifetimes of a million years \citep{enoch2008}.  If we assume that our cold cores are mainly starless, our lifetime estimates exceed this half million year estimates by an order of magnitude.  On the other hand, if we assume that our cold cores are a combination of starless, and Class 0/I type protostars, then this estimates is shorter than ours by a factor of a few, though, given the considerable uncertainties, perhaps marginally consistent with what we see. 
 
It should be noted that many of these surveys have been made in regions of known star formation.  Since they typically depend on the observed ratios of less evolved cores to more evolved cores, selection on the existence of evolved objects could bias the estimates low.  BLAST has shown that there can be considerable populations of cold cores far from any evidence of heating by later stages of star formation.  For example, region\,C in Figure~\ref{fig:velafullcolor} contains 169 cold cores, with a mean temperature of 13\,K, including one of over 80\,M$_\odot$, and with no cores over the 20\,K temperature expected for envelopes being blown off from Class II protostars \citep{molinari2008}, other than three associated with RCW-36.

An even more striking contrast is with high mass star formation lifetime estimates found from an analysis of Cygnus X\@. \citet{motte2007} find lifetimes for starless and protostellar cores of less than $10^4$\,years for cores with $M>40\mbox{\,M}_\odot$, based on the ratio of the early cores to more evolved types within the map.  

Though we do not currently have a complete census of all O stars in our region, we can evaluate the consistency of our lifetimes by applying a similar reasoning to our data.  Taking the definition of high mass star forming cores from \citet{motte2007} as cores with $M>40\mbox{\,M}_\odot$, we integrate our fit to the mass function in Figure~\ref{fig:hist_M}.  We predict 7.8 high mass cold cores in our map (we actually see 8).  Integrating our estimated normalized star formation rate from our region given in Equation (\ref{eqn:rv}) we predict that there should have been 9 stars with $M>40\epsilon\,\mbox{M}_\odot = 12\,\mbox{M}_\odot$ formed in our region over the past 2\,million years, which is consistent with the visual impression of hot (blue) spots in Figure~\ref{fig:velafullcolor}.  Averaging over very large areas, we find that the lifetimes for cold cores is on the order of millions of years, and far in excess of the free-fall times, even for the high mass tail of our distribution.

The existence of even these small numbers of high mass cores in Vela is strong evidence for a long lifetime.  However, our lifetime estimate could be compromised because we may have misidentified these objects as high-mass star forming cores; for example, these objects could be the result of the blending of many low mass cores, or our mass estimates could be significantly off.  But our resolution is close to what \citet{motte2007} use to define an object as a core rather than a clump, and our sources are all well below the clump scale.  Our masses are calibrated to C$^{18}$O masses, and an underestimated distance (more likely than an overestimate) would produce underestimated source masses, making the discrepancy even larger.  On the other hand, since Cygnus X was chosen as a region undergoing significant amounts of high mass star formation (which is a very rare event), lifetime estimates based on statistics of the region could be strongly biased toward shorter values - the region is being observed at a very special time.

\section{Discussion}

Since the measured lifetimes for the intermediate mass cores significantly exceed the free-fall time, and since the cores are too cold and dense to form stable thermal pressure supported objects, this long lifetime is only possible if there is some other sort of support.

Supersonic turbulence has been proposed as one possible mechanism for preventing the collapse of clouds and clumps, consistent with the observation that the C$^{18}$O line widths on the clump scale are in the range of 1--2~\kms.  In fact, if turbulence were great enough, our cores could be transient objects, and not stable at all.  However, as turbulence tends to decay on a sound crossing time --- a time similar to the free fall time --- it is challenging to invoke turbulence as a mechanism to support cores for millions of years.  However, outflows from embedded protostars can serve to maintain turbulence and extend the embedded phase \citep{matzner2007}, and for these intermediate mass cores, significant low mass stellar activity could be present without strongly warming up the cores.  A solar luminosity per solar mass only warms up the dust in the core to $\approx18$\,K.

There is some evidence that cold cores, at least at low masses, are not highly turbulent.  NH$_3$ and $\mbox{N}_2\mbox{H}^+$ line widths in low mass cores in the Pipe Nebula and Ophiuchus have been measured to be $\approx0.15$\,\kms\  \citep{rathborne2008, andre2007} --- insufficient to prevent collapse for most of our cores.  This has led to the conclusion that these cores are short lived \citep{andre2007}.  If the intermediate mass cores that we observe have line widths comparable to these low mass cores, then turbulence would be inadequate to provide support for the times required.

Magnetic support can also provide a mechanism for producing starless core lifetimes much longer than the free-fall time.  The 4\,Myr lifetimes we find here are consistent with initial flux to mass ratios near critical \citep{tassis2004} --- though with stronger fields the cores could last much longer than this.  Whether the fields within cores are actually strong enough remains a point of contention \citep{crutcher2008, mouschovias2008}.  The relatively long cold core lifetimes we find, however, do lend weight to magnetic support playing an important role in slowing intermediate mass core collapse.

Whether the long lifetimes we find are mediated by turbulence generated by some ongoing process within the core, or by magnetic fields, can not been determined by these data alone.  Followup observations of these cores in NH$_3$ or  $\mbox{N}_2\mbox{H}^+$ could be used to determine if there is enough turbulent motion to inhibit the collapse.  \citet{olmi2009} combine BLAST data with IR observations  carried out by {\it Spitzer} of a subset of this map to allow a more sensitive search for embedded stars and protostars than is possible with \emph{MSX} or \emph{IRAS}.  Additionally, submillimeter polarization measurements by the upcoming BLASTpol \citep{marsden2008} mission will help to determine if the magnetic fields in cores have the coherence to larger scales expected for models of magnetic field mediated star formation.

We acknowledge the support of NASA through grant numbers NAG5-12785, NAG5-13301, and NNGO-6GI11G, the NSF Office of Polar Programs, the Canadian Space Agency, the Natural Sciences and Engineering Research Council (NSERC) of Canada, and the UK Science and Technology Facilities Council (STFC).   CBN acknowledges support from the Canadian Institute for Advanced Research.   This research made use of WestGrid computing resources. 

\bibliography{ms}

\LongTables
\label{sec:fulltable}
\begin{deluxetable*}{cccrrrc} %{ccrrrrr}
\tabletypesize{\scriptsize}
\tablewidth{0pt}
\tablecaption{BLAST Vela Sources Measured Properties
\label{tab:measuredtable}
}
\tablehead{
\colhead{Source name} &
\colhead{$l$} &
\colhead{$b$} &
\colhead{$F_{250}$} &
%\colhead{$\sigma_{250}$} &
\colhead{$F_{350}$} &
%\colhead{$\sigma_{350}$} &
\colhead{$F_{500}$} &
%\colhead{$\sigma_{500}$} &
\colhead{FWHM$_{250}$} \\
\colhead{} &
\colhead{(\degree)} &
\colhead{(\degree)} &
\colhead{(Jy)} &
\colhead{(Jy)} &
\colhead{(Jy)} &
%\colhead{(Jy)} &
%\colhead{(Jy)} &
\colhead{(\arcsec)} }
\startdata
      BLAST J083512-404528&259.9986& -0.1588&     8.9$\pm$     1.2&     6.1$\pm$     0.9&     2.8$\pm$     0.6& 50.3\\ %        1208
      BLAST J083524-404027&259.9553& -0.0774&    66.3$\pm$     9.6&    24.0$\pm$     6.7&     7.1$\pm$     4.4& 72.5\\ %         140
      BLAST J083528-404129&259.9762& -0.0780&    42.0$\pm$     6.6&    13.3$\pm$     4.8&     3.3$\pm$     3.4& 72.2\\ %        1000
      BLAST J083530-403819&259.9378& -0.0414&   280.5$\pm$    28.8&   118.9$\pm$    15.2&    43.4$\pm$     7.0& 42.1\\ %          20
      BLAST J083547-420809&261.1670& -0.8979&    62.3$\pm$     6.4&    30.6$\pm$     4.0&    13.8$\pm$     2.2& 48.1\\ %         154
      BLAST J083604-420049&261.1010& -0.7823&     9.3$\pm$     1.6&     4.1$\pm$     1.3&     1.3$\pm$     1.0& 61.8\\ %         810
      BLAST J083606-410224&260.3271& -0.1914&     9.4$\pm$     1.2&     7.0$\pm$     1.0&     3.6$\pm$     0.6& 54.6\\ %        1277
      BLAST J083607-404415&260.0865& -0.0083&    22.3$\pm$     3.3&    12.6$\pm$     2.5&     5.1$\pm$     1.6& 71.7\\
\enddata
\tablecomments{Source properties as determined in Section~\ref{sec:fits}.  Flux densities are quoted at
 precisely 250, 350 and 500~\micron\ using SED fits to obtain
 color-corrections for the band-averaged flux densities.
}
  %(Section~\ref{sec:coldsed}). The quoted statistical uncertainties
  %are determined from Monte Carlo simulations, and do not include
 %calibration uncertainties.
%}
%\tablenotetext{a}{These sources are located on ripples in the
%  deconvolved map and the BLAST colors are considered unreliable as a
%  result.}  \tablenotetext{b}{V07 is believed to lie in the outer
%  Galaxy.}  \tablenotetext{c}{These sources are associated with a
%  molecular cloud in the Perseus arm.}  \tablenotetext{d}{Also has a
%  comparable component at $-$15\,km\,s$^{-1}$.}
%\tablenotetext{e}{Also has a comparable component at
%  23\,km\,s$^{-1}$.}
\end{deluxetable*}

\begin{deluxetable*}{crrrc} %{ccrrrrr}
\tabletypesize{\scriptsize}
\tablewidth{0pt}
\tablecaption{BLAST Vela Sources Derived Properties
\label{tab:derivedtable}
}
\tablehead{
\colhead{Source name} &
\colhead{Temperature} &
\colhead{Mass} &
\colhead{Luminosity} &
%\colhead{$\sigma_{250}$} &
\colhead{Size} \\
%\colhead{$\sigma_{350}$} &
%\colhead{$F_{500}$} &
%\colhead{$\sigma_{500}$} &
%\colhead{FWHM$_{250}$} \\
\colhead{} &
\colhead{(K)} &
\colhead{(\Msun)} &
%\colhead{(Jy)} &
%\colhead{(Jy)} &
%\colhead{(Jy)} &
%\colhead{(Jy)} &
\colhead{(\Lsun)} &
\colhead{(pc)} }
\startdata
BLAST J083512-404528 & 13.0 $\pm ^{1.8}_{1.2}$ & 4.3 $\pm ^{1.9}_{1.6}$ & 2.2 $\pm ^{1.4}_{0.9}$ & 0.12 $\pm$ 0.01\\ 
BLAST J083524-404027 & 33.0 $\pm ^{42.5}_{13.3}$ & 1.8 $\pm ^{3.4}_{1.5}$ & 249 $\pm ^{9282}_{153}$ & 0.21 $\pm$ 0.02\\ 
BLAST J083528-404129 & 51.6 $\pm ^{30.4}_{30.8}$ & 0.5 $\pm ^{1.9}_{0.3}$ & 993 $\pm ^{9790}_{1248}$ & 0.21 $\pm$ 0.02\\ 
BLAST J083530-403819 & 22.0 $\pm ^{3.0}_{1.3}$ & 20.4 $\pm ^{6.2}_{7.5}$ & 246 $\pm ^{211}_{67}$ & 0.08 $\pm$ 0.00\\ 
BLAST J083547-420809 & 17.1 $\pm ^{2.0}_{0.9}$ & 9.7 $\pm ^{2.5}_{3.3}$ & 26.1 $\pm ^{14.0}_{7.1}$ & 0.11 $\pm$ 0.00\\ 
BLAST J083604-420049 & 22.2 $\pm ^{19.6}_{9.2}$ & 0.7 $\pm ^{1.6}_{0.6}$ & 8 $\pm ^{660}_{63}$ & 0.17 $\pm$ 0.02\\ 
BLAST J083606-410224 & 12.0 $\pm ^{1.3}_{0.8}$ & 6.6 $\pm ^{2.1}_{2.4}$ & 2.0 $\pm ^{0.8}_{0.6}$ & 0.14 $\pm$ 0.01\\ 
BLAST J083607-404415 & 15.6 $\pm ^{4.4}_{3.1}$ & 5.0 $\pm ^{4.6}_{2.7}$ & 7.7 $\pm ^{51.8}_{2.0}$ & 0.21 $\pm$ 0.02\\
\enddata
\tablecomments{Source properties determined from SED fits as described
in Sections ~\ref{sec:sed} and \ref{sec:sourceProperties}.  The fits assume a 
dust emissivity index, $\beta=$\,2.0, a dust mass absorption coefficient, 
$\kappa r=0.16\mbox{cm}^2\mbox{g}^{-1}$, and a distance of 700\,pc.  
The size is from the deconvolved FWHM.
}
  %(Section~\ref{sec:coldsed}). The quoted statistical uncertainties
  %are determined from Monte Carlo simulations, and do not include
 %calibration uncertainties.

%\tablenotetext{a}{These sources are located on ripples in the
%  deconvolved map and the BLAST colors are considered unreliable as a
%  result.}  \tablenotetext{b}{V07 is believed to lie in the outer
%  Galaxy.}  \tablenotetext{c}{These sources are associated with a
%  molecular cloud in the Perseus arm.}  \tablenotetext{d}{Also has a
%  comparable component at $-$15\,km\,s$^{-1}$.}
%\tablenotetext{e}{Also has a comparable component at
%  23\,km\,s$^{-1}$.}
\end{deluxetable*}

\end{document}